# Applying a Chemical Structure Teaching Method in the Pharmaceutical Analysis Curriculum to Improve Student Engagement and Learning[1]

Hui Zheng[a,*], Binjing Hu[a], Qiang Sun[b], Jun Cao[a] and Fangmin Liu[a]

[a]College of Material, Chemistry and Chemical Engineering, Hangzhou Normal University, Hangzhou 311121, China

[b]ARC Centre of Excellence for Nanoscale BioPhotonics (CNBP), School of Science, RMIT University, Melbourne, VIC 3001, Australia

**Abstract:** Pharmaceutical analysis, as the core curriculum of chemistry, chemical engineering and pharmaceutical engineering, contains broad and in-depth knowledge that leads to massive learning & teaching loads. There are more than 100 analytical methods of medicines in this course. As such, this subject is a big challenge for both students and lecturers. A novel chemical structure teaching (CST) method was developed based on our long-term teaching experience to cope with these challenges. It has been shown in practice that this CST method can significantly unload the stress of students and lecturers simultaneously. The survey about the improvement of students' interests was carried out and listed in the form of questionnaire. The outcome of CST also indicates that it can help them to form abilities of critical and logical thinking as independent learners, motivate them to discuss with their peers and lecturers, and eventually improve average grades. Furthermore, CST can be beneficial for lecturers who instruct other relevant curriculum in chemical or pharmaceutical engineering to improve the teaching outcome, such as organic chemistry, spectrum analysis, pharmaceutical synthesis and medicinal chemistry. This CST model can also help students cultivate lifelong learning ability as active learners and habit from the cognitive perspective view.

## 1. Introduction

The development of skills such as problem-solving, critical-thinking and lifelong active learning is very important for their future career and living.[1] Nevertheless, the inefficient capability cultivating in traditional teacher-oriented style is still dominated in most universities,[2] in which students are merely required to receive the information from lecturers who then have to try to memorize it in a mechanical way. As such, the attitudes and self-concepts of students were pushed aside to the secondary place, which can normally leads to lack of motivation and engagement in learning.[3] In particular, for students whose major is chemistry, chemical engineering or pharmacy, many courses contain diverse, complex and in-depth knowledge such as organic chemistry and medicinal chemistry. It is more difficult to lecture and learn effectively *via* that traditional lecturing procedure. Given the deficiencies and shortcomings of the conventional teaching process, novel and explicit teaching methods need to be developed to enable students to engage and be active learners, and eventually to develop themselves to be proficient and independent problem-solvers.[4]

In 1987, Chickering and Gamson[5] published a seminal paper that identifies seven universal practices in undergraduate education: 1.encouraging student-faculty contact; 2.encouraging cooperation among students; 3.encouraging active learning; 4.giving prompt feedback; 5.emphasizing time on task; 6.communicating high expectations; 7.respecting diverse talents and ways of learning.

From then on, many novel education methods have been implemented in chemical and pharmacy related courses based on these seven principles, such as incorporating drug discovery stories in a pharmaceutical chemistry class[6] can greatly boost student-faculty interactions. Romero RM et al [7] have applied problem-based learning (PBL) method in pharmaceutics course to assess its effectiveness in comparison with didactic approach. Results showed that students who took the PBL lecture scored significantly higher than those in didactic lecture. Malcolm L. H. Green et al [8] developed covalent bond classification (CBC) method to help students understand relationships between molecules more comprehensively, which improves their ability to conceptualize and master the chemical properties of the elements.

Pharmaceutical analysis, one of the core courses in pharmacy related majors and one of the required subjects for licensed pharmacist examination, is a discipline that is needed to meet the demand of overall quality control in drug R&D and manufacturing. It covers chemical and instrumental analysis methods ranging from synthetic medicine to natural medicine, biological medicine and their agents.[9] For students to master this course, both rich knowledge in chemistry and pharmacy and good experimental skills are indispensable. So pharmaceutical analysis is usually set in junior year to meet the requirement of combining cross-disciplinary knowledge and skills that students have learnt in previous semesters that build solid foundation for their future employment or research work, such as medicine quality control and analysis.

The pharmaceutical analysis course usually includes pandect (4 chapters) and sub-pandect (12 chapters).[10] The former covers the overview of fundamental analysis principles, methods and common technologies, underlining the similarities and characteristics between drug analytical approaches. In general, pandect comprises drug standard, identification test, impurity test, assay, biopharmaceutical analysis and validation of analytical methods.[10] As for the sub-pandect, it includes the knowledge on identification, impurity limit test and assay with more than 100 typical medicines, such as aromatic acids non-steroid anti-inflammatory drugs, *p*-aminobenzoate acid ester, etc. There are up to 12 categories in total. In addition, the analytical methods of bulk drugs and preparations from Chinese pharmacopoeia (abbreviated as ChP 2015 edition, the same below) [11] are also covered in this course. All of these diverse, complicated and in-depth contents definitely present barrier to students' productive learning.

From our practical teaching experience, we find that students tend to lose initiatives, become aimless and feel anxious when facing so many complex chemical structures and analytical methods. Nevertheless, we notice that it is always much easier and more interesting for them to learn pandect than the subsequent sub-pandect chapters. The problem might be the lack of clear and uniform logic to instruct the whole teaching and learning process. As a result, there is no surprise that students are more facile to engage with pandect materials, which have clear logic and are mainly based on what students have learnt previously, relative to how they feel when learning sub-pandect.

Given the teaching and learning dilemma in pharmaceutical analysis, we firmly believe that a new education method is needed to enhance students' learning interests and efficiency. Herein, to deal with the above problem arising from teaching and studying process, we developed a unique education method named as "chemical structure teaching (CST)" after continuously exploring for many years. As suggested by the name, CST mainly focuses on the chemical structures of different drugs. With this method, students will easily and actively learn how to deduce possible properties and analytic methods according to specific functional groups, namely, to link what they've learnt before in organic chemistry and analytic chemistry with pharmaceutical analysis.

There are some papers examining the efficiency of chemistry combined into pharmacy teaching. Structurally Based Therapeutic Evaluation (SBTE) is such a valuable and practical teaching approach in medicinal chemistry introduced by Alsharif et al[12]. It mainly emphasizes the relevance of chemistry to the practice of pharmacy and the structurally based therapeutic evaluation. Therein, students are required to identify the chemical/structural basis for the pharmacological and therapeutic action of drugs, then to analyze and explain why a drug works. Specifically, CST is to translate chemical and structural understanding into predicted chemical properties, and then speculate its possible analysis method.

According to the connectivism theory by George Siemens[13], in the digital era where knowledge and information increase exponentially, to know is in fact to be connected. That's to say, learning process is analog to weaving web, and namely when we are learning something new, we are actually trying to put the new knowledge into our existing networks by adding new nodes. Therefore, the more connective the knowledge network is built, the more holistic our understanding will be. Connectivism theory has already been incorporated in medical education and exhibited positive influences.[14] The same principle can also be applied in pharmaceutical analysis. Repeated employment of CST is beneficial to the efficiency of memory retrieval in related courses.

In summary, CST can free students from heavy memory burdens, enable students to hone skills in synthesizing information, thinking logically, and analyzing problems independently. Also, it can be utilized in other curriculums, such as organic chemistry, medicinal chemistry and spectrum analysis.

## 2. The discovery of CST

After analyzing and summarizing the contents of pharmaceutical analysis, it is found that all chapters in the textbook[10] are arranged following the following logic: identification test, impurity test, assay, biopharmaceutical analysis and analysis method validation. Inspired by the fact that pharmacological activities are always determined by the chemical structure, we can conclude that the learning materials of identification test, impurity test and assay are also closely related to drugs' chemical structures. As such, we developed this novel and effective CST method based on our long-term teaching experience which has been practically applied in our pharmaceutical analysis course. The key idea of CST is to focus on the chemical structures of medicines due to the basic principle "properties are determined by structures". For instance, the concrete identification, impurity test and assay procedures are likely to be inferred by logical deduction. The whole process of CST is designed as "observing structures→speculating properties→determining analysis methods→verifying with textbook and the ChP" as shown in Fig. 1. This process has been proven conducive to both students and teachers. As such, when learning a new drug, students can actively speculate possible analysis methods based on drug's chemical structure, and then verify correctness of inferred methods with ChP and textbook.

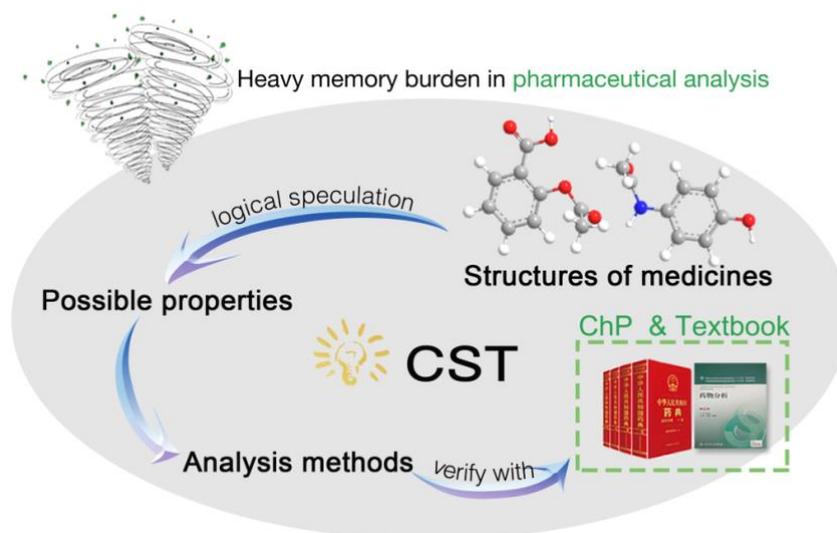

**Fig.1 Graphical representation of CST method**

The creative teaching mode can provide meaningful references for other related courses such as organic chemistry, medicinal chemistry, spectrum analysis, pharmaceutical synthesis, to just mention a few, as displayed in Fig.2.

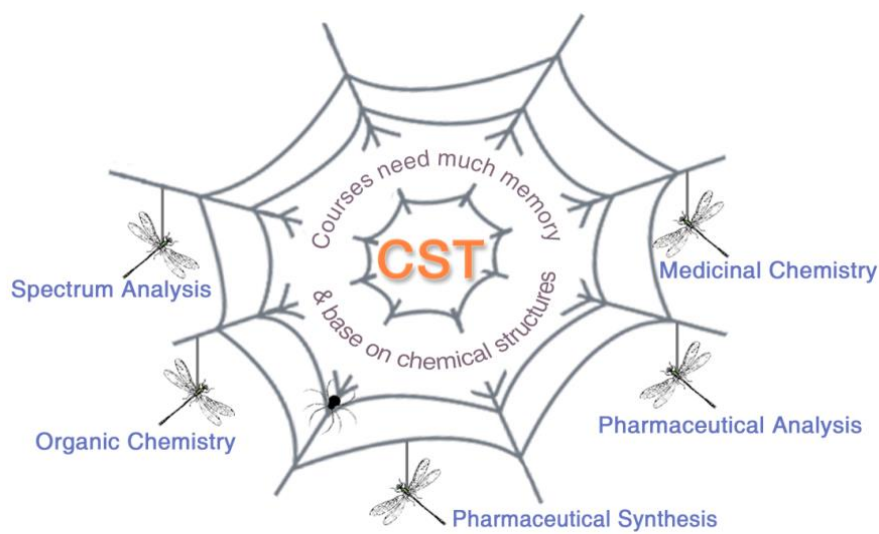

**Fig.2 The possible extended curriculum of structure-based teaching method**

## 3. The application of CST

After identifying the fundamental logic of pharmaceutical analysis course, we designed some teaching steps to implement this method. When lecturing pandect, we generally introduce the CST method to make students have an initial impression on CST. When stepping into lecturing sub-pandect, we choose some chapters as examples to demonstrate CST in detail. All students are encouraged to take part in the design process for that chapters, which will be beneficial to train their ability of critical thinking, oral expression and logical reasoning. The typical application of CST in pharmaceutical analysis is detailed as follows.

First, we guide students to be familiar with the basic process of CST. They are trained to deduce possible physical and chemical properties according to the elementary structures of medicine，then further to explore possible impurities and analysis methods with preliminary assumptions. We then let students summarize analytical methods from the aspect of identification, impurity test and assay, respectively, which are verified using the corresponding knowledge mentioned in textbook and the ChP to strengthen the impression of key points. At last, we guide students to draw a frame diagram for each drug to visualize and methodize piecemeal knowledge points on the chemical structures of drug. Following this process, students can review what they have studied and build a strong link between structure, properties of drugs and analytical methods.

Take the CST expression of Quinolones as an example. As demonstrated in Fig.3, we can see that Quinolones have relatively large molecule weights and carboxyl group which can form hydrogen bond easily. On the basis of these observations, we can deduce that quinolones might be solid and can be identified by melting point. Due to the weak alkalinity of tertiary amine group, quinolones can be converted into products in brown, red, purple or blue when co-heating with malonic acid in anhydride. Nonaqueous titration is also applicable in quantitative analysis of quinolones. The carboxyl groups suggest that quinolones can be identified or assayed by acid-base reactions, and decarboxylic product might be the source of impurity. In addition, we can make use of the reverse synthesis principle in organic chemistry to speculate possible impurities, such as the unreacted raw materials or intermediates or solvents.

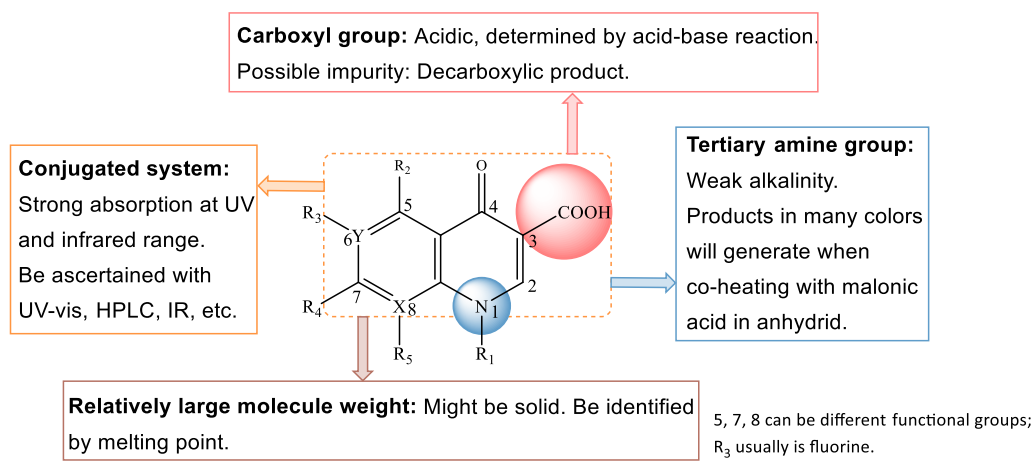

Fig.3 The expression of Quinolones structure-based teaching method

Practice is crucial in learning. After introducing the key points of CST and taking several chapters as examples, students will practice CST by themselves. We organize a few lectures where students and lecturer swap their roles. Students are then divided into groups and design the lectures with their peers. This has been proven to be an effective way in cognitive education and quality education[1b]. During this learner-based teaching process, instructors can observe how students master the learning materials, give timely feedback, and help them consolidate the key points in pharmaceutical analysis and resonant with correlated curriculums.

Also, reasoning method and case study method along with inquiry teaching, heuristic teaching, discussion teaching, situational teaching and other teaching methods can be applied synergistically to enhance students' interests and to encourage them to engage with learning materials. Some examples of implementing these methods in our practical teaching are illustrated as the followings.

1. Carry out case study. Since the drugs introduced in pharmaceutical analysis are already widely used in society, we can discuss a piece of negative news of one drug in class. Then, the lecturer can guide students to make a connection between the learning materials with the society news by thinking and analyzing what can cause a negative effect on drugs in society. For example, the transportation and storage of drugs which fail to follow GSP may lead to unwanted impurities, which could ultimately lead to dreadful tragedies. Case study is an effective method to highlight question guide, to increase learning interests to a great extent and to let students explore and solve problems actively.

2. According to the course goals, teachers can combine deductive and inductive teaching methods in lectures. For instance, in some lectures, teacher can act as an organizer, an instructor and a listener or a student, to encourage students to discuss and learn actively.

3. Teachers can carry out frame teaching in pharmaceutical analysis. Following the requirement of ChP, students need to make identification, do impurity test and assay with the chemical structures of drugs. This can be demonstrated clearly by drawing frame diagrams to visualize key points. It can also be used in organic chemistry, analytical chemistry, spectrum analysis, medicinal chemistry, etc.

## 4. Results and discussion

### 4.1 Interest survey

After completing the lectures to undergraduate majored in pharmaceutical engineering in 2016-2018, a questionnaire entitled 'Improvement of pharmaceutical analysis learning quality in CST method' was distributed to them so as to obtain their opinions about the CST method. The total number of students who have completed and returned the questionnaires is 98. And the questionnaire comprises five questions to evaluate the learning load after the mastery of CST method, learning interest and also, the adoptability in other related chemical courses. All of these five questions are ranging from strongly disagree to strongly agree, and the results is shown in Fig.4.

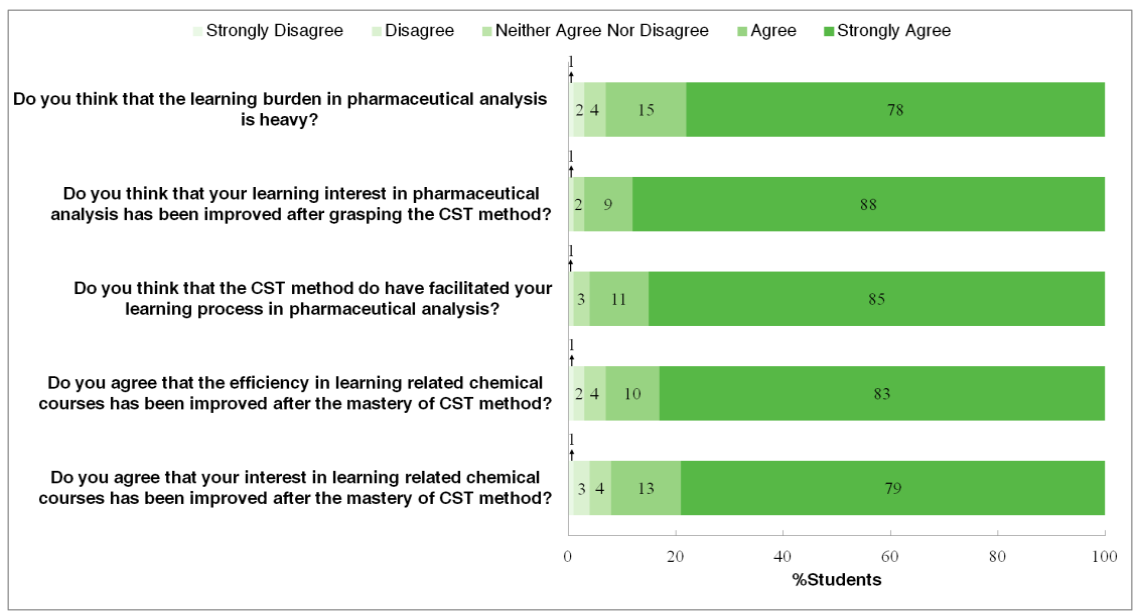

**Fig.4 Results of the questionnaires distributed to the students major in pharmaceutical engineering (The number of students that answered the questionnaire is 98)**

As demonstrated in Fig.4, at least 92% of the students answered positively (agreed or strongly agreed) to these five questions, and only no more than 8% students gave negative (disagreed or strongly disagreed) answers in general, which shows that the majority of students found themselves beneficial in learning pharmaceutical analysis with CST method.

To be specific, there are 93% of the students believed that the learning burden in pharmaceutical analysis is real heavy. However, after the implementation of CST method, nearly 96% of the students found their interest aroused in a more facilitated studying process. It also revealed that more than 92% of the students have found their efficiency and interest in other related chemical courses enhanced as well. These results reveal that the initial efforts in CST method implementation are successfully paid off as the majority of students gave positive answers.

## 4.2 Teaching effects

The effectiveness of CST in pharmaceutical analysis is investigated by some general factors, such as the average scores, the ability to synthesize information, logical thinking, and analysis of problems independent of students. These results can give some information about the CST teaching effects.

The students' average scores of final exams from 2013 to 2018 were counted, and the results are shown in Fig.5. Within those 6 years, CST method was used into pharmaceutical analysis teaching in 2016-2018, while the traditional script-oriented teaching was used in 2013-2015. From Fig.5, the average scores have an increase after the implementation of the CST method, which is positive evidence to support the CST teaching effect.

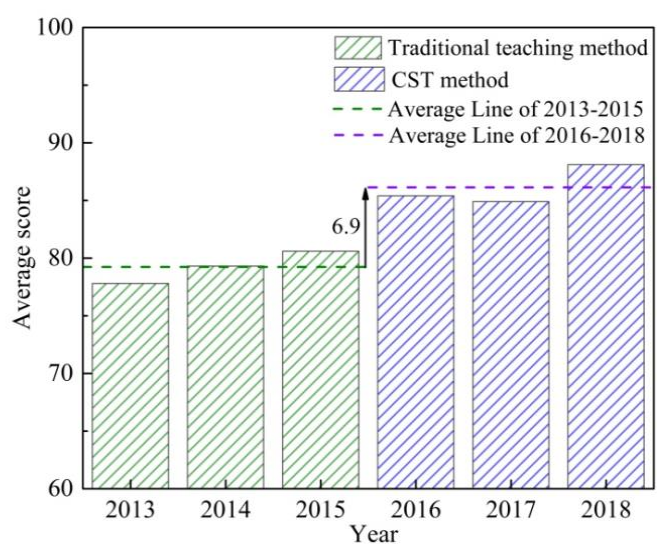

**Fig.5 Average scores of pharmaceutical analysis in 2013-2018 (The number of students from 2013 to 2018 is 36, 34, 35, 34, 34, 30, respectively)**

Similar to the survey of learning quality about the CST method, the questionnaires regarding synthesizing information, thinking logically, and analyzing problems independently of student ability were employed to evaluate the teaching effects. Also, all 98 participant questionnaires were returned, and the results were summarized in Fig. 5. For the synthesizing information ability, 98% students thought the CST method was helpful to promote this ability as well as the ability of cultivating independent analytical problem-solving. There were 96% students who thought the CST method could promote their logical thinking ability. So, it was found that most students thought the CST method had a positive teaching effect.

In addition to the improved performances on exams and the relevant abilities, CST also exhibited a great capacity for helping students increase good habits and qualities which are indispensable for their future career and life. First, through the CST process, students are trained to focus on the fundamental chemical structures of medicines, namely, they can see through the phenomenon to study the essence. Thereafter, students can gradually figure out the relationships between those seemingly different and diverse courses, which is analog to assemble jigsaw puzzles, to map out a bigger picture.[15] In addition, the ability to identify properties and analyze methods from structures, in other words, to decipher the information contained within the medicines' structures, will help students to think logically, increase confidence and motivate learning interests. Also, while employing CST, teachers can guide students to form the connections between new pieces of knowledge and old knowledge network, which will eventually change their original modes of thinking. As George Siemens wrote in *Connectivism: A learning theory for the digital age*,[16] the pipe is more important than the content within the pipe. That is, today, study is to link continuously instead of to always construct, which can ultimately weave a broader and broader web of knowledge, so that the implementation of CST can accelerate students' acquirement of new information while strengthening the old in the meantime. They will also tend to engage with new learning materials actively and to become life-long active learners. Last but not the least, CST can be used in other curriculums with similar characteristics such as organic chemistry.

Due to the heavy-duty memorization-based learning materials in the courses of pharmacy majors, it is very much needed to use a robust method which can help students to quickly master learning materials more efficiently. CST has been proven effective to ease the burden in pharmaceutical analysis teaching and learning. Also, in the process of CST, students can get timely feedback when verifying their anticipated ideas with reference to the requirements of ChP. Together with other teaching methods, such as group learning, study efficiency can be further improved.[17]

## 5. Conclusion

In this paper, CST method has been demonstrated on the basis of the problems encountered when lecturing pharmaceutical analysis. CST was developed based on the drug's chemical structure, and then has been made a series of logical inferences as far as different analytical methods. The CST method can help students and teachers ease the learning and teaching burden which is resulted from the complex and scattered knowledge points existing in pharmaceutical analysis. Our experience and practice have proven that the CST can not only let students master key points more easily, more autonomous and more efficient, but also can help them engage with learning materials more, learn more actively, interact with faculty and their peers more. Given the similar characteristics, CST can also be applied in other courses. For example, in organic chemistry, CST can help to speculate chemical properties of structures. In spectrum analysis, CST can be used to demonstrate how to determine different chromatographic behaviors and spectral characters. We believe that CST can be popularized amongst the chemistry courses whose knowledge systems are diverse, memorization-based and focusing on the relationships between chemical structures and properties. In conclusion, CST is an effective teaching mode and method suitable to learners and lecturers who have chemistry background to learn and teach pharmaceutical analysis or other related chemistry curriculums.

**Acknowledgement**

Q.S. would like to acknowledge the support from the Australian Research Council through a Centre of Excellence grant (CE140100003) and a Discovery Early Career Researcher Award (DE150100169). Also, we are grateful to Zhejiang Province Higher Education Teaching Reform project and The Second batch of Collaborative Projects of Ministry of Education of China (201802066080).